\documentclass[9pt,twocolumn,twoside]{osajnl}
\usepackage{color}

\journal{ol} 

\setboolean{shortarticle}{true} 

\title{A 30 GHz electro-optic frequency comb spanning 300 THz in the near infrared and visible}

\author[1,2,*]{Andrew J. Metcalf}
\author[1,2]{Connor D. Fredrick}
\author[1,3]{Ryan C. Terrien}
\author[1,2]{Scott B. Papp}
\author[1,2,*]{Scott A. Diddams}
{\setlength{\fboxsep}{0pt}%
 \setlength{\fboxrule}{0pt}%

\affil[1]{Time and Frequency Division, National Institute of Standards and Technology, 325 Broadway, Boulder, CO 80305, USA}
\affil[2]{Department of Physics, University of Colorado, 2000 Colorado Ave., Boulder, Colorado 80309, USA}
\affil[3]{Carleton College, Northfield, MN  USA}

\affil[*]{Corresponding authors: andrew.metcalf@colorado.edu, scott.diddams@nist.gov}

\begin{abstract}
Beginning with a continuous wave laser at 1064 nm, we generate a 30 GHz electro-optic frequency comb which contains 100 lines spanning 3 THz. The initial comb is subsequently amplified, spectrally broadened in normal dispersion photonic crystal fiber, and then temporally compressed to provide 74 fs pulses with average power of up to 2.6 W.  When launched into a second photonic crystal fiber with anomalous dispersion, a supercontinuum spanning 800-1350 nm is generated. Second harmonic generation allows for extension of the 30 GHz comb into the visible, yielding greater than 300 THz of total spectral bandwidth.  Such a broad bandwidth, high repetition rate comb is a compelling source for astronomical spectrograph calibration.
\end{abstract}

\setboolean{displaycopyright}{false}

\begin{document}

\maketitle

Broad bandwidth optical frequency combs with mode spacing in the range of 10-100 GHz are unique sources for applications that benefit from access to and control of individual comb modes. Specific examples include arbitrary waveform generation \cite{weiner:2010}, high-speed communications \cite{kippenberg:2014}, photonic signal processing \cite{delfyett:2006,metcalf:2016}, spectroscopy \cite{heinecke:2009}, and astronomical spectrograph calibration \cite{murphy:2007,quinlan:2012,steinmetz:2008,braje:2008,ycas:2012,li:2008}. In this high-frequency regime, electro-optic (EO) based combs have gained considerable attention due to their robust operation, high power per mode, ease of external RF synchronization, and flexibility in tuning the center frequency and mode spacing independently \cite{kourogi:1993,imai:1998,metcalf:2013,torres:2014,aubourg:2015,wang:2010,ishizawa:2013}.

The majority of EO comb schemes have been demonstrated in the communication band at 1.5 $\mu$m where they make use of mature lithium-niobate modulator technology and benefit from flexibility in fiber based dispersion control. EO comb sources typically generate a few nanometers of bandwidth supporting the synthesis of picosecond pulses, however, many applications require more spectral coverage. Nonlinear spectral broadening in optical fiber and nonlinear waveguides has been successfully used to extend the spectrum of $10-30$ GHz EO combs centered at 1.5 $\mu$m to greater than octave bandwidth \cite{beha:2017,carlson:2017}. And at the same time, powerful techniques have been employed to minimize the impact of multiplicative microwave noise \cite{beha:2017,carlson:2017} that had previously been a primary limitation in broad bandwidth EO comb generation \cite{imai:1998}.

\begin{figure}[b!]
\centering
\includegraphics[width=\linewidth]{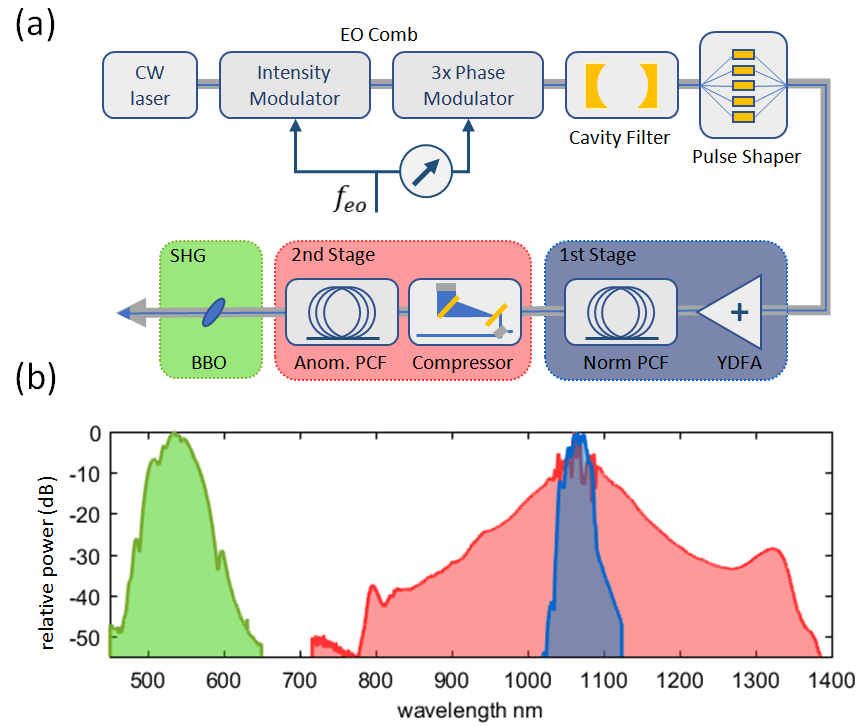}
\caption{(a) Layout of the 1$\mu$m 30 GHz EO comb experiment. (b) Summary of spectra output from the first stage (blue) of spectral broadening in normal dispersion photonic crystal fiber (PCF), the second stage (red) of broadening in anomalous dispersion PCF, and finally second harmonic generation (green) in beta-barium borate (BBO). All spectral amplitudes are normalized to their respective peak values.}
\label{fig:1}
\end{figure}

There has also been interest in developing EO combs in the 1-$\mu$m region \cite{wang:2010, chapman:2013,prantil:2013,aubourg:2015}.  Here the recent progress in integrated lithium-niobate modulator technology, coupled to the ease of synchronization with an external frequency reference could have particular advantages for microscopy \cite{wang:2010} and the generation of photoelectron bunches in accelerators \cite{Li:2018}. Moreover, as repetition rates increase above 10 GHz the average power required to reach high pulse energy for nonlinear spectral broadening and harmonic generation can become a limiting factor. In the 1$\mu$m region, this can be addressed by ytterbium (Yb) power amplifiers that offer unparalleled average power and the mature catalog of supporting components designed to handle such powers. 

In this paper, we describe a 1$\mu$m EO comb source that provides pulses as short as 74 fs at a 30 GHz rate with 2.6 W of average power. For robustness and long-term stability, the core parts of the system are implemented in polarization maintaining (PM) fiber. This consists of a base EO comb generated with four cascaded modulators, followed by dispersion management, amplification, and controlled spectral broadening in normal dispersion photonic crystal fiber (PCF). Subsequent spectral broadening of this unique ultrashort pulse source in an anomalous dispersion PCF with a parabolic dispersion profile leads to a spectrum spanning 800-1350 nm.  Throughout the nonlinear broadening, high contrast is maintained on the 30 GHz comb lines across 150 THz. This robust source constitutes the core infrastructure to serve as the calibrator of the Habitable Zone Planet Finder (HPF), which is a high-resolution spectrograph designed to find exoplanets around M dwarf stars \cite{mahadevan:2012, metcalf:2019} .  Finally, we demonstrate the capability of extending the spectral coverage of the 30 GHz comb into the visible through second harmonic generation.  This provides approximately 300 THz of total spectral coverage in the visible and near infrared.

\begin{figure}[b!]
\centering
\includegraphics[width=\linewidth]{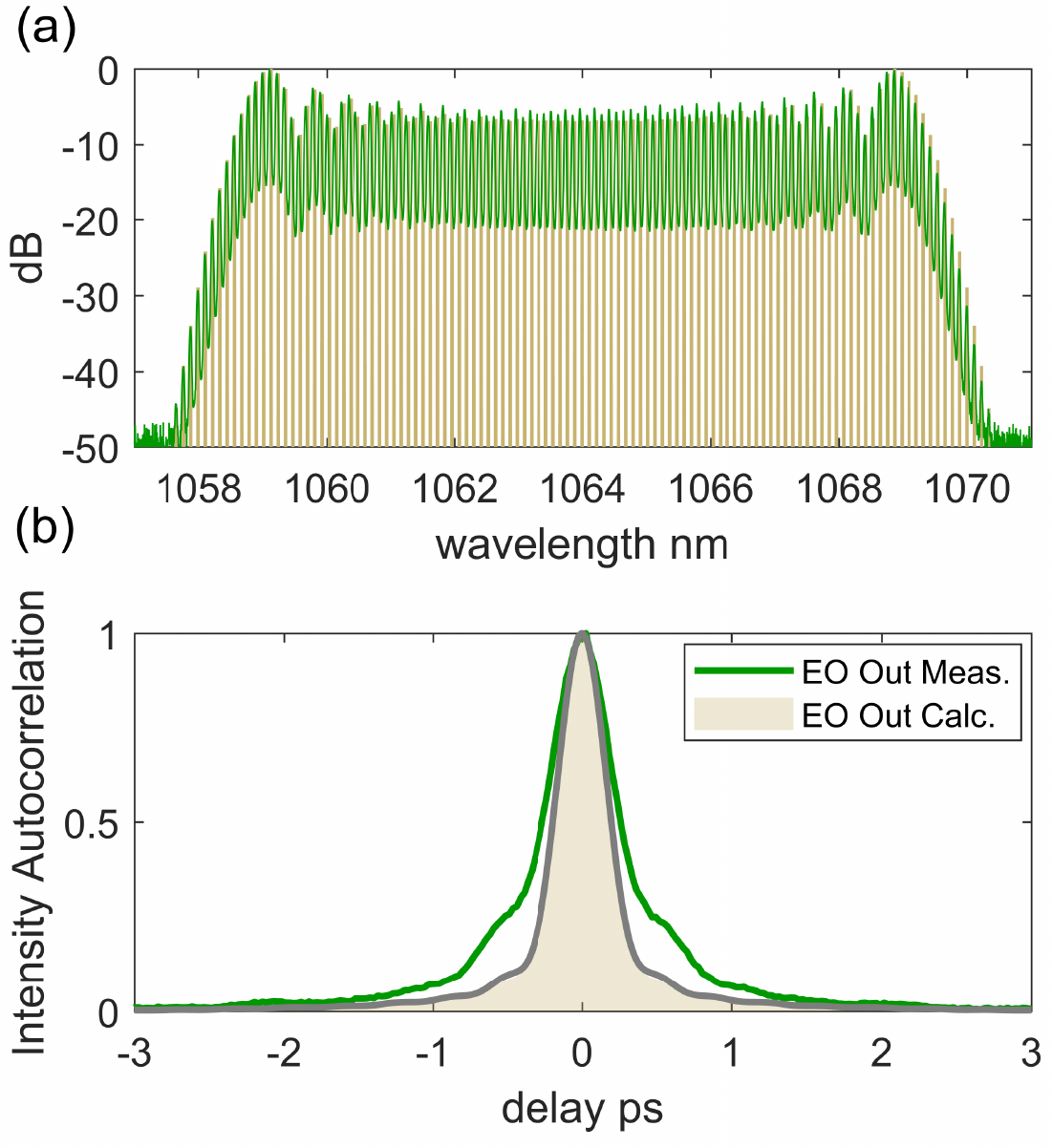}
\caption{Output of EO comb, (a) spectra- simulated (tan) and measured (green), (b) time domain intensity autocorrelations - measured (green) and calculated band-limited using recorded spectrum and assuming a flat spectral phase (tan).}
\label{fig:2}
\end{figure}
Figure \ref{fig:1} shows a schematic of our setup and provides a summary of the output spectra measured after each stage. Our EO comb generation scheme is similar to that in \cite{metcalf:2013} and consists of a continuous wave (CW) laser centered at 1064 nm, three lithium-niobate phase modulators (PMs), and one intensity modulator (IM) arranged in series.  Each modulator is driven by the same 30 GHz RF source. The IM DC bias is adjusted to carve out a flat top pulse from the CW laser and the PMs are aligned in time using RF phase shifters to apply a strong linear frequency chirp across the pulse.  The PM V$_{\pi}$ values rages from 2.5-3 V at 1 GHz and are each driven by a narrow band 3 W RF amplifier.  Each PM produces around 33 comb lines resulting in a comb spanning $\sim$3 THz (see Fig.\ref{fig:2}(a)). The linear chirp applied by the PMs leads to a nearly quadratic spectral phase which can be well compensated using single mode fiber alone. Figure \ref{fig:2}(b) shows the measured intensity autocorrelation (AC) trace after propagation through single mode fiber overlaid with the band-limited AC which was calculated using the measured spectrum and assuming a flat spectral phase. Using a deconvolution factor of 1.65 the compressed pulse width is $\sim$330 fs.

\begin{figure}[t!]
\centering
\fbox{\includegraphics[width=\linewidth]{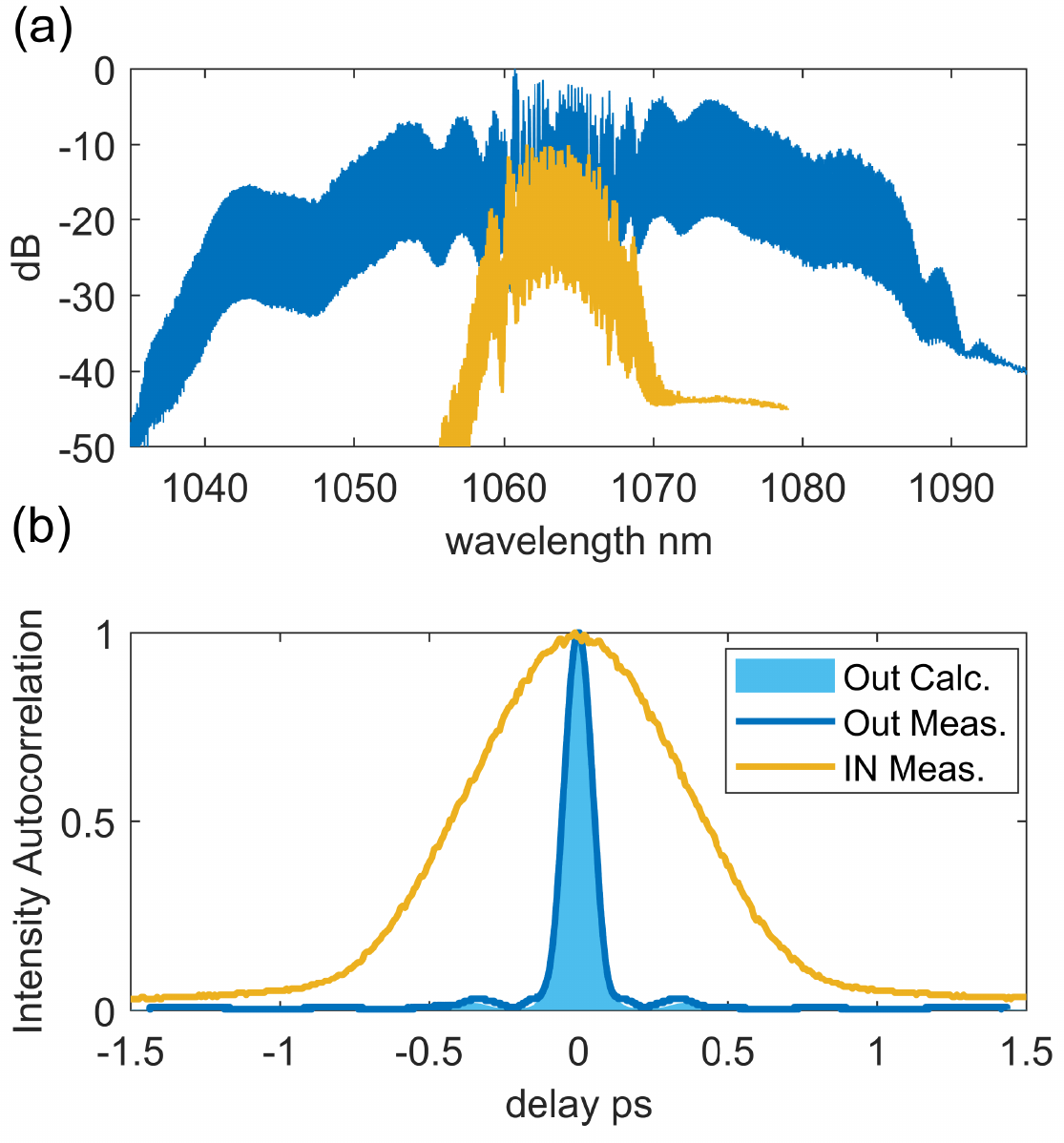}}
\caption{Data from the first stage of broadening in Normal dispersion PCF.  (a) Measured spectra at the input (yellow) and output of PCF1 (blue). (b) Intensity autocorrelation traces at the input of PCF1 (yellow) and output after passing through a grating compressor (blue-line). Calculated band-limited autocorrelation using measured spectra at output of PCF1 and assuming a flat phase (blue-shaded).  The FWHM of measured and calculated autocorrelations are $\sim$100 fs.}
\label{fig:3}
\end{figure}
We pass the output of the EO comb though a PM-fiber coupled 30 GHz Fabry-Perot optical cavity with finesse of approximately 600 in order to suppress thermal noise before the subsequent amplification and nonlinear broadening stages \cite{beha:2017}. This filter cavity has its length actively stabilized to the 30 GHz frequency comb. Additionally, we employ a programmable pulse shaper with 27 GHz resolution (Finisar 1000A) to fine tune the spectral phase.  Although the EO comb can be compressed using single mode fiber alone--as demonstrated above--the amount of dispersion accumulated in fiber components between the output of the EO comb and the broadening stages exceeds the amount required to compress the pulse. The pulse shaper allows us to keep the full comb generation setup, including compression and amplification, contained in all polarization maintaining fiber-coupled components. Additionally, the pulse shaper enables us to make fine adjustments to the pulse chirp in real-time while monitoring the broadened output of the PCF. 

Although our EO comb produces relatively short 330 fs pulses directly, it has been shown that pulse durations greater than 200 fs can negatively affect supercontinuum generation in anomalous dispersion fiber \cite{dudley:2006,tamura:2000}.  To circumvent this issue we divided our spectral broadening into two stages, denoted as 1st Stage and 2nd Stage in Fig.\ref{fig:1}(a).  The first stage utilizes a 10 m piece of PM normal dispersion PCF with a nonlinear coefficient of 37 W/km (NKT NL-1050-NEG-PM-FUD \cite{Liu:15}). Normal dispersion broadening helps to suppress modulation instability and retain the coherence of the broadened spectra \cite{tamura:2000,huang:2008}.  Figure \ref{fig:3}(a) shows the input (yellow) and output (blue) spectra of the first broadening stage when the input optical power is set at 4.1 W. The PCF coupling efficiency was $\sim$78\% which left 3.2 W of optical power and sufficient bandwidth to support a pulse duration of $\sim$70 fs. The ends of the PCF were capped with coreless fiber to expand the mode and terminated in SMA connectors to improve coupling efficiency and provide stability when working at high average powers. After exiting the PCF the broadened spectra is sent through a grating compressor which shortens the pules to $\sim$74 fs at their full width half max (FWHM).  Figure \ref{fig:3}(b) shows the measured AC traces at the input of the PCF (yellow) and after the grating compressor (blue). Also included is the calculated band-limited trace (light blue) which was calculated using the measured spectrum at the output of the PCF and assuming a flat spectral phase.  The output power from the grating compressor was measured at $\sim$2.6 W which corresponds to a peak power of $\sim$1.2 kW. 

\begin{figure}[t!]
\centering
\fbox{\includegraphics[width=\linewidth]{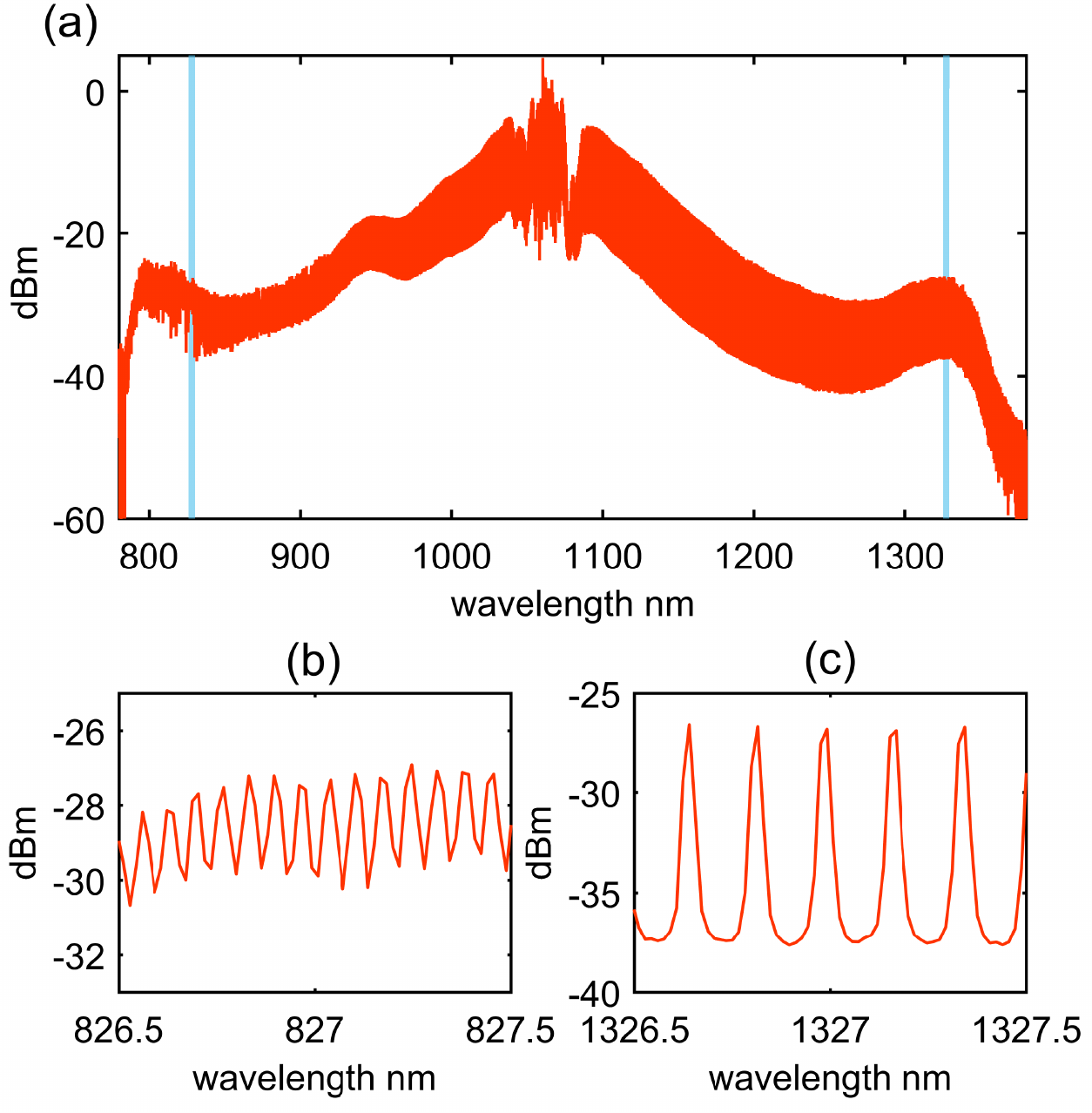}}
\caption{Output spectra from the second stage of broadening in anomalous dispersion PCF. (a) Full spectrum taken at 0.02nm resolution with a Optical Spectrum Analyzer (OSA). (b,c) Zoomed in spectra corresponding to blue shaded regions in (a). We note, the jagged nature of the zoomed traces are a representitive of the OSA.}
\label{fig:4}
\end{figure}

After exiting the grating compressor the band-limited pulse train is directed to the second stage of broadening which consists of a 0.5 m piece of anomalous dispersion PCF (NKT ZERO-1050-2).  The output of Stage2 is then coupled into a 7 m piece of SM980 fiber and recorded with an optical spectrum analyzer (OSA) at 0.02 nm resolution. The measured spectrum is plotted in Fig.\ref{fig:4}(a) and exhibits a smooth spectral envelope which extends more than 150 THz, covering the wavelength range from 800-1350 nm.   Figure \ref{fig:4}(b and c) provide zoomed-in traces of the resolved 30 GHz comb modes at the short and long wavelength regions of the spectrum, respectively. The extinction on the short wavelength side is limited by the resolution of the OSA. The comb power measured in free space directly after the PCF was 1.8 W which corresponds to more than 1 $\mu$W/mode at the lowest part of the spectrum.

\begin{figure}[t!]
\centering
\fbox{\includegraphics[width=\linewidth]{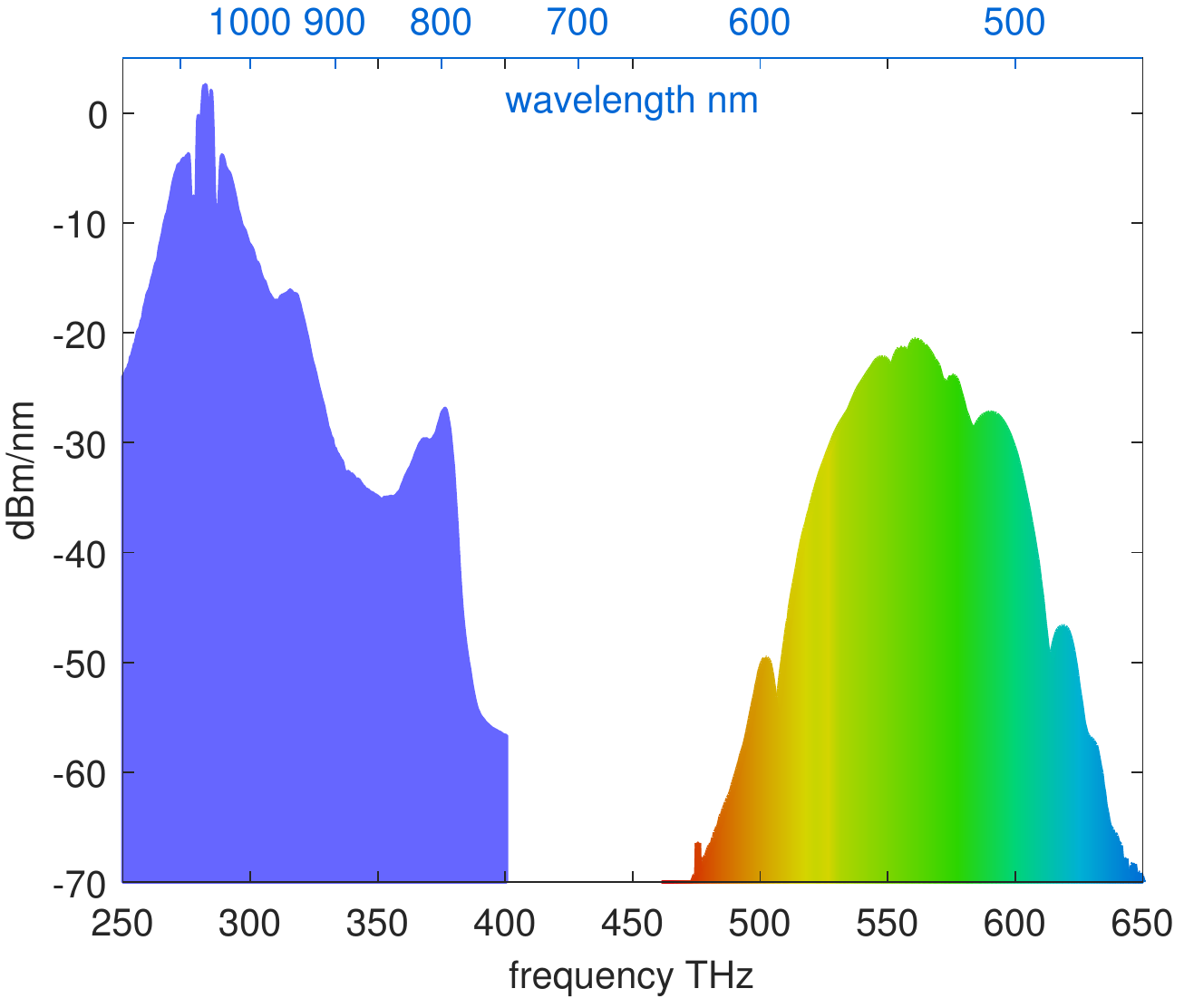}}
\caption{Measured spectra at the input (purple) and output (colored) of the SHG stage.  The frequency-doubled spectra was isolated using a filter to remove the fundamental pump centered at 1064 nm and then recorded using the MaxHold function on the OSA while rotating the orientation of the crystal with respect to the incident beam.}
\label{fig:5}
\end{figure}

The high average powers can enable extension of the broadened 1 $\mu$m comb to the visible.  Taking the free space output of Stage2 and focusing it into a thin BBO crystal generates a 30 GHz comb at twice the center frequency through a combination of SHG and sum frequency generation (SFG).  After the crystal the fundamental 1 $\mu$m comb is filtered out and the remaining visible light is coupled into a large core fiber for measurement with the OSA.  The OSA was set on \emph{MaxHold} as we gradually adjusted the orientation of the crystal to provide phase matching over the largest possible bandwidth.  The output spectrum measured at 1 nm resolution is shown in Fig.\ref{fig:5} alongside a portion of the original spectrum (purple) for comparison.  The visible comb spans 150 THz in total with over 70 THz in a 10 dB bandwidth.  The total power in the visible comb is $\sim$1 mW.

In summary, we have generated a 30 GHz comb source centered at 1064 nm, compressed its pulses to \textless{}100 fs in duration, and extended its bandwidth to cover more than 150 THz in the NIR with over 1 $\mu$W/mode. Combs with mode spacing greater than 10 GHz and large spectral coverage are ideal candidates for calibrating astronomical spectrographs.  Further, we have demonstrated the ease by which our 1 $\mu$m NIR comb can be frequency doubled to produce a broadband frequency comb in the visible--a region of the spectrum that has traditionally been challenging to reach with frequency comb technology having mode-spacings in the 10+ GHz range.

\bigskip
This work is a contribution of the National Institute of Standards and Technology (NIST) and not subject to copyright in the US.  Mention of specific products is for scientific information only and does not constitute an endorsement by NIST.   
\smallskip

The authors acknowledge support from the National Science Foundation (NSF) AST-1310875 and the National Institute of Standards and Technology. The authors thank J. Jennings and F. Quinlan for technical support.

\bigskip


\bibliographyfullrefs{sample}
 

\ifthenelse{\equal{\journalref}{aop}}{%
\section*{Author Biographies}
\begingroup
\setlength\intextsep{0pt}
\begin{minipage}[t][6.3cm][t]{1.0\textwidth} 
  \begin{wrapfigure}{L}{0.25\textwidth}
    \includegraphics[width=0.25\textwidth]{john_smith.eps}
  \end{wrapfigure}
  \noindent
  {\bfseries John Smith} received his BSc (Mathematics) in 2000 from The University of Maryland. His research interests include lasers and optics.
\end{minipage}
\begin{minipage}{1.0\textwidth}
  \begin{wrapfigure}{L}{0.25\textwidth}
    \includegraphics[width=0.25\textwidth]{alice_smith.eps}
  \end{wrapfigure}
  \noindent
  {\bfseries Alice Smith} also received her BSc (Mathematics) in 2000 from The University of Maryland. Her research interests also include lasers and optics.
\end{minipage}
\endgroup
}{}

\end{document}